# Fuzzy approach on modelling cyber attacks patterns on data transfer in industrial control systems

Emil Pricop
Petroleum-Gas University of Ploiesti
Ploiesti, Romania
emil.pricop@upg-ploiesti.ro

Sanda Florentina Mihalache
Petroleum-Gas University of Ploiesti
Ploiesti, Romania
sfrancu@upg-ploiesti.ro

*Abstract* – **Cybersecurity of industrial control system is a very complex and challenging research topic, due to the integration of these systems in national critical infrastructures. The control systems are now interconnected in industrial networks and frequently to the Internet. In this context they are becoming targets of various cyber attacks conducted by malicious people such as hackers, script kiddies, industrial spies and even foreign armies and intelligence agencies. In this paper the authors propose a way to model the most frequent attacker profiles and to estimate the success rate of an attack conducted in given conditions. The authors use a fuzzy approach for generating attacker profiles based on attacker attributes such as knowledge, technical resources and motivation. The attack success rate is obtained by using another fuzzy inference system that analyzes the attacker profile and system intrinsic characteristics.**

## I. INTRODUCTION

Control systems are very complex technical structures composed not only by sensors, classical controllers and actuators, but also integrating devices with significant processing power and networking capabilities.

Nowadays each industrial system integrates various automated control systems that are interconnected on industrial networks at the plant level and connected to the Internet for remote operation and monitoring. These systems are frequently the target of various cyber attacks conducted by malicious persons from script kiddies and disgruntled employees to cyber terrorists and industrial spies. Since many of these industrial systems are components of national critical infrastructure, their security is a big challenge for industry, government and academia.

The authors model the attacker profiles and estimate the attack success rate using a fuzzy approach. Before defining the fuzzy inference rules, we have to clarify some concepts related to control.

Identification of threats and vulnerabilities is needed, when analyzing the control systems security [5]. Threats are synonym to attackers, even if they are identified as people or malicious hardware/software components. Threats are hard to characterize, since they have a high indetermination degree and can not be prevented. On the other side vulnerabilities are breaches which can be exploited by a threat. Vulnerabilities can be detected and controlled, the system administrator being able to protect the system against known issues [2].

The threats analyzed in this paper are attackers. Each attacker has a set of attributes such as knowledge of the target system, technical resources, motivation to conduct an attack against the system. All these characteristics allowed us to make some attacker profiles, as presented in [1] and in Section III of this paper.

In the next sections the authors provide a detailed methodology for modelling the attacker profiles and for estimating the attack success rate by using a fuzzy inference approach.

## II. FUZZY INFERENCE SYSTEMS

The fuzzy inference systems (FIS) are an appropriate solution to describe into a linguistic manner the behavior of a process [3]. The process behavior is usually described by a mathematical or linguistic model. The FIS approach is more suitable than conventional modelling methods in the case of nonlinear processes, if information about process nonlinearities is available [4].

The architecture of a general fuzzy inference system is presented in the figure 1. A fuzzy inference system is formed by 5 functional units:

- the fuzzification unit: at this phase (the premise stage) the crisp value of the input is transformed into a set of matching degrees to membership functions describing the fuzzified input;

- the defuzzification unit: at this phase (the consequence stage) the fuzzy set resulted from rule inference mechanism is transformed into a crisp value;

- the database unit: the storage unit for membership functions parameters for both input and output fuzzy variables (model's tuning parameters);

- the rule base unit: the storage unit for the IF-THEN rules that describe system behavior (Mamdani-M or Takagi-Sugeno TS type);

- the decision block: the inference mechanism is applied to the existing rules.

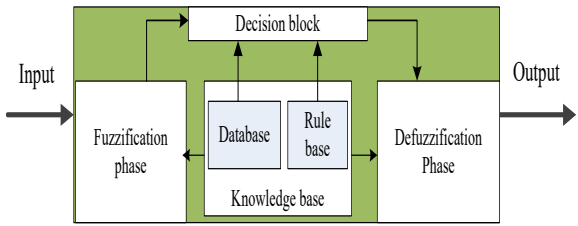

Figure 1. The architecture of a fuzzy inference system

The mathematical modelling of the process behavior usually begins with finding the proper membership functions for input and output variables (first set of tuning parameters of the model – the proper shape for membership functions: triangular, bell shaped, Gaussian, trapezoidal et al.). After choosing the proper shape, the tuning parameters for chosen membership functions are found using trial and error methods (second tuning parameters of the model). This phase is followed by the construction of the fuzzy rules (M or TS type). The IF-THEN rules are the result of observations and experience of an expert in process field. The last phase is applying the operations of inference to the fuzzy rules and then establishing the crisp output value and the modelling result.

In the following sections is proposed a fuzzy approach on modelling the cyber attacks patterns on industrial control systems. In authors opinion the cyber attacks can be described with attributes such as attacker profile, protection degree of attacked unit, system vulnerabilities and restore costs (impact factor).

### III. MODELLING THE ATTACKER PROFILE

The cyber attacks that happen in industrial control systems are conducted by different types of attackers. The "perpetrators" go from un-skilled "script kiddie" to experts "cyber warrior".

The authors experience suggests that an attacker profile can be associated to an attacker score, indicating the ability of an attacker to successfully conduct an attack.

Each attacker can be characterized by at least three parameters: knowledge, technical resources and motivation [1]. These parameters are presented in the following paragraphs.

Knowledge is an intrinsic attacker characteristic, that indicates his technical skills that can be used to initiate and conduct an attack. Also this parameter can show the awareness on the target system architecture and the existent countermeasures. A script kiddie, an Internet user that finds an exploit and tries to use it, has a very low knowledge on the target system; meanwhile an industrial spy or a disgruntled employee is aware of the system specifications, security measures and has much more chances to exploit a specific vulnerability than the common Internet user.

Technical resources parameter is an attacker profile attribute that indicates the hardware and software resources that can be used in order to deploy a specific attack type.

Motivation is an intrinsic attacker attribute that indicate the determination to conduct an attack on the system. Motivation is given by financial win, reputation, revenge. This parameter is hard to quantify and is very dynamic.

These are the three inputs for the proposed FIS. The output is the attacker profile score. If three membership functions are used as granularity of model description for each input, it results in 27 fuzzy rules that describe attacker's profile. Increasing the number of membership functions (MF) for each input generates a significant growth of fuzzy rules number (MF number of inputs), with little consequence on the quality of the model versus computational effort. This is the consequence of the importance in fuzzy modelling of the completeness, continuity and consistency of fuzzy rule base. In Table 1 are presented the most important features of proposed membership functions (trimf-triangular, trapmf - trapezoidal) for proposed inputs.

The output is described by 5 triangular membership functions as depicted in figure 2.

TABLE I. INPUTS DESCRIPTION

| Input | Membership functions | | |
|---|---|---|---|
| | *Small* | *Medium* | *Big* |
| Resources | trapfm [-0.225 -0.025 0.1 0.5] | trimf [0.3 0.6 0.9] | trapfm [0.7 0.9 1.06 1.26] |
| Knowledge | trimf [-0.4 0 0.5] | trimf [0 0.5 1] | trimf [0.5 1 1.4] |
| Motivation | trapfm [-0.45 -0.05 0.1 0.4] | trimf [0.2 0.5 0.8] | trapfm [0.6 0.95 1.05 1.45] |

The shapes and parameters for membership functions resulted from trial and error methods and authors experience in fuzzy modeling. The experience in attacker behavior is described by the 27 fuzzy rules (Mamdani type).

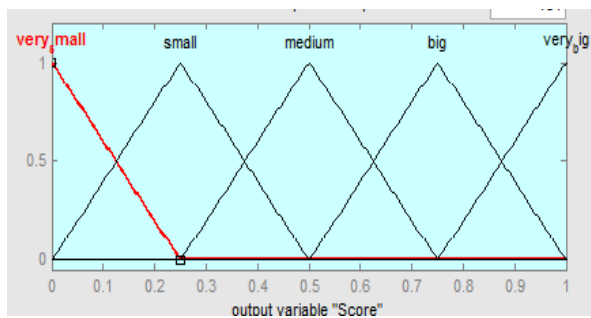

Figure 2. Output of 5 triangular membership functions

Attacker profiles can be generated by various combinations of the three parameters presented in the previous paragraphs. Each combination is possible, but in the literature and taking into account the authors experience we can define some well-established profiles:

- Script kiddie is an unexperienced and unskilled attacker that found a security exploit and tries to use it in order to obtain reputation. This type of attacker has rarely success in targeting protected systems, even if they initiate the majority of attacks.

- Hacker is a skilled attacker that initiates attacks for gaining reputation in his community. He may have good technical resource. Generally, he has little knowledge about the target system architecture.

- Disgruntled employee has a very high level of knowledge about the system. May initiate very dangerous attacks from the interior of the industrial network. His motivation is medium and decreases over time. The main scope is to revenge against the employee.

- Terrorists are now employing skilled people with a decent resources level available. They can work as individuals or as organized groups. Their motivation is for financial gain and for spreading terror and clutter in the society. They are targeting mainly critical systems.

- Industrial spy tries to get access to confidential data or to sabotage the competition. They have very high knowledge and technical resources, being specialists in the field, and also a good financial gain motivation.

- Cyber warrior is the most dangerous attacker. This attacker has the highest levels of knowledge, resources and motivation. They can initiate and sustain distributed attacks against many targets simultaneously. Their objective is to break the functioning or to destroy critical infrastructures. This kind of attackers are sustained by enemy countries in the context of electronic and information warfare.

Table II offers some particular cases of attacker profiles and the corresponding main active rule.

TABLE II. PARTICULAR CASES OF KNOWN ATTACKERS

| Attacker type | Main active rule | | | | |
|---|---|---|---|---|---|
| | *No* | *Res.* | *Know.* | *Motiv.* | *Score* |
| Script kiddie | 1 | small | small | small | very small |
| Hacker | 4 | small | medium | small | small |
| Disgruntled employee | 8 | small | big | medium | medium |
| Terrorist | 14 | medium | medium | medium | medium |
| Industrial spy | 26 | big | big | medium | very big |
| Cyber warrior | 27 | big | big | big | very big |

The attacker profile is a function of three inputs and one output. The resulted FIS shows that the major impact of determining the attacker score is associated to knowledge, followed by resources and motivation.

For example, if motivation is medium, the attacker scores depend on resources and knowledge as in figure 3.

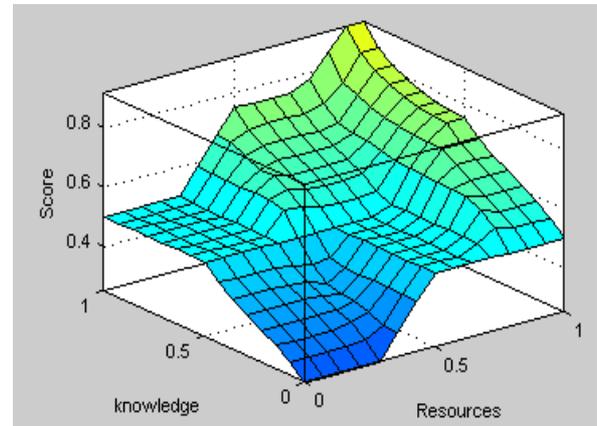

Figure 3. Attacker score in case of medium motivation

Furthermore, if resources levels are small, like in the case of disgruntled employee, the attacker score dependence on knowledge is presented in figure 4, with a nonlinear saturation shape.

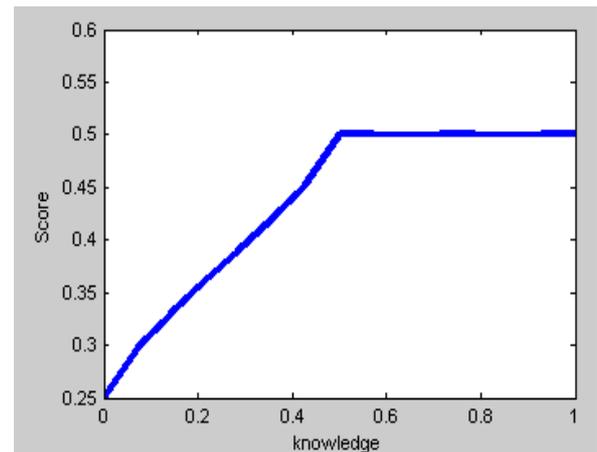

Figure 4. Digruntled employee - attacker profile score

If resources level is big, like in the case of industrial spy, the attacker score dependence on knowledge is presented in figure 5.

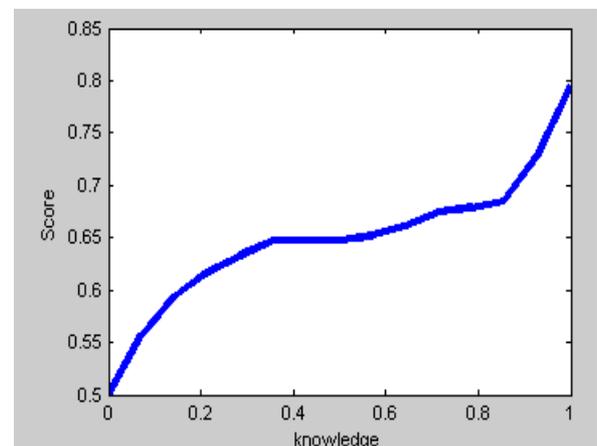

Figure 5. Industrial spy case – attacker profile score

If resources are medium, like in the case of terrorists, the dependence of attacker score is presented in figure 6 (again saturation shape).

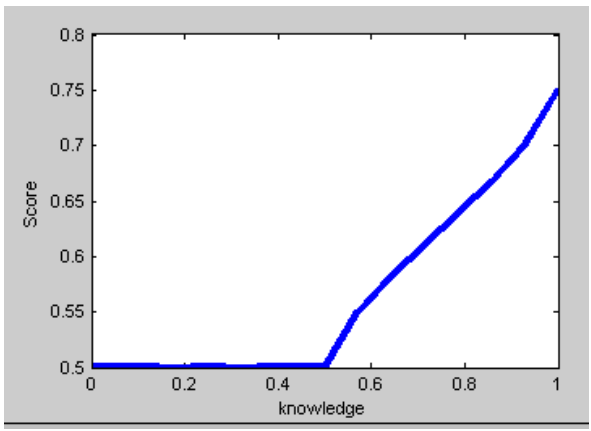

Figure 6. Terrorist – profile attacker score

It is obvious that the attacker profile is proportional in a way with the knowledge. The dependence between profile score and knowledge are in the presented cases nonlinear increasing functions. Of particular interest are the extreme cases of script kiddie and cyber warrior.

In the case of script kiddie or hacker, the motivation is small, and the resulted surface of costs depending on knowledge and resources is presented in figure 7.

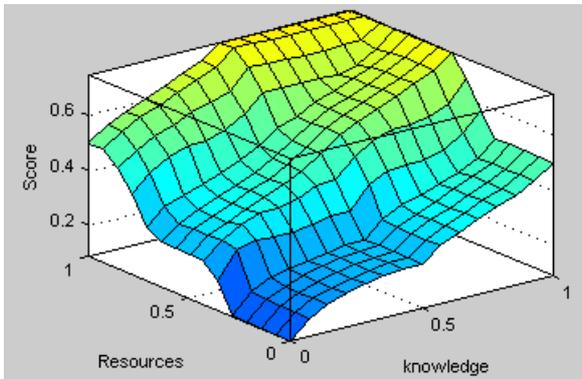

Figure 7. Attacker profile score in case of small motivation

If resources are small (the case of script kiddie), the attacker score depending on knowledge are presented in figure 8 and if resources are medium (the case of hacker) the scores are presented in figure 9.

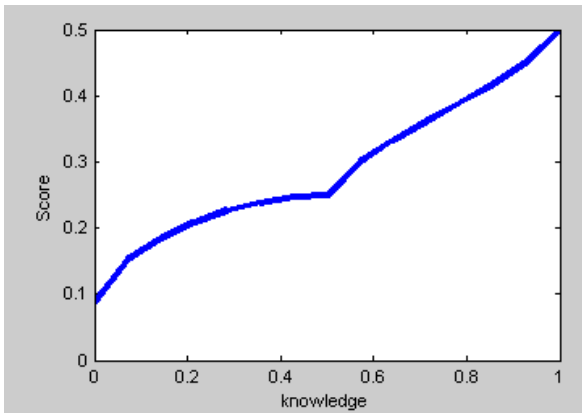

Figure 8. Script kiddie - attacker profile score

Both characteristics show almost linear behavior, only in the case of script kiddie there are two distinct portions with different proportional gains.

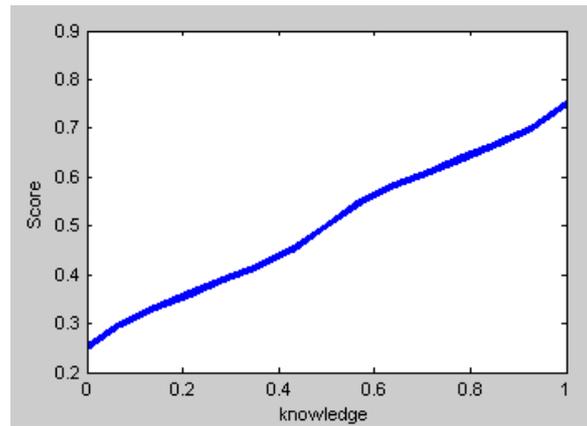

Figure 9. Hacker case - attacker profile score

The extreme case of big motivation determines the surface presented in figure 10.

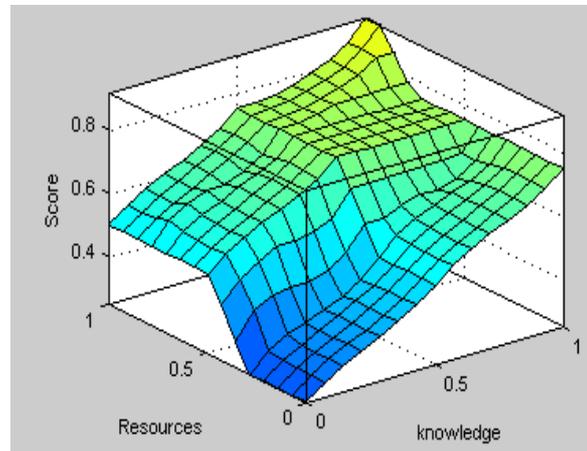

Figure 10. Attacker profile score in care of big motivation

If resources are big, like in the case of cyber warrior knowledge has the influence on attacker score as presented in figure 11.

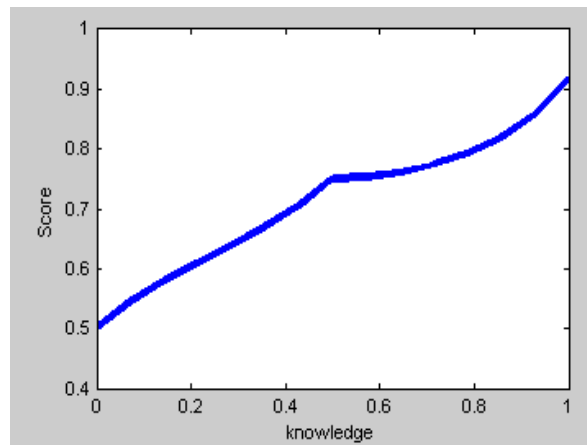

Figure 11. The cyber warrior - attacker profile score

The above simulation results made in MATLAB show the importance of knowledge factor in establishing the scores associated with attacker profile.

The influence of resources is less important than knowledge as seen from figure 12 (motivation medium, knowledge big). This is the case of disgruntled employee (small resources) and industrial spy (big resources).

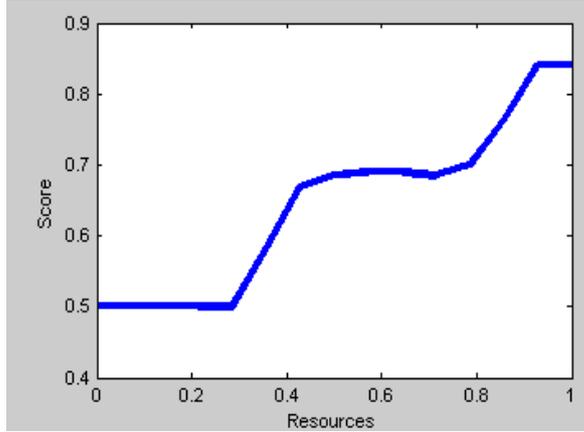

Figure 12. Saturation in case of disgruntled employee and industrial spy

## IV. MODELLING THE ATTACK SUCCESS RATE

After modeling the attacker profile, the authors propose a fuzzy model for the attack success rate.

We define the *attack success rate* as the ability of the attacker to interrupt the functioning of the target system. This rate is mostly influenced by four parameters: the attacker profile score previously established, the system protection level, the system vulnerabilities and the restore cost.

The system protection level is an intrinsic system characteristic related to the number and the quality of countermeasures taken in order to secure the system.

The system vulnerabilities indicator shows the level of the known vulnerabilities that are not mitigated by the countermeasures already taken.

The restore cost is an important factor that indicates both the time and the financial resources needed to re-establish the correct system functioning in case of an attack.

The proposed inputs parameters are presented in table 3.

TABLE III. INPUTS DESCRIPTION

| Input | Membership functions | | |
|---|---|---|---|
| | *Small* | *Medium* | *Big* |
| Attacker profile | trimf [-0.5 0 0.5] | trimf [0 0.5 1] | trimf [0.5 1 1.5] |
| Protection level | trimf [-0.4 0 0.3] | trimf [0.1 0.4 0.7] | trimf [0.4 1 1.4] |
| Vulnerabilities | trimf [-0.4 0 0.4] | trimf [0.1 0.5 0.8] | trimf [0.6 1 1.4] |
| Restore cost | trimf [-0.4 0 0.4] | trimf [0.1 0.5 0.8] | trimf [0.7 1 1.4] |

Each input is described by three membership function with triangular shape. This yields to 81 fuzzy rules Mamdami type. The output (successrate) has 5 associated membership functions like in figure 13.

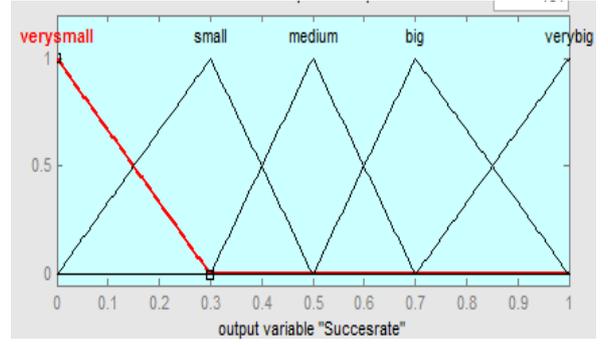

Figure 13. Output successrate representation

The fuzzy rules corresponding to the four inputs describe cyber attack pattern in industrial control systems. In Table 4 is presented a selection of the 81 resulted rules (VS-very small, S-small, M-medium, B-big, VB-very big).

TABLE IV. SELECTION OF PROPOSED FUZZY RULES

| No. | Main active rule | | | | |
|---|---|---|---|---|---|
| | *Attacker Profile* | *System protection* | *System vulnerabilities* | *Restore cost* | *Success rate* |
| 1 | S | S | S | S | VS |
| 2 | S | S | S | M | S |
| 6 | S | S | M | B | M |
| 10 | S | M | S | S | VS |
| 15 | S | M | M | B | S |
| 20 | S | B | S | M | VS |
| 27 | S | B | B | B | S |
| 35 | M | S | B | M | B |
| 40 | M | M | M | S | S |
| 45 | M | M | B | B | B |
| 60 | B | S | M | B | VB |
| 70 | B | M | B | S | B |
| 81 | B | B | B | B | VB |

The rules are constructed according to authors' experience. For example, rule 27 represents a less likely to happen attack, upon a highly protected target with big restore costs by an unskilled attacker so the success rate is small. Rule 60 specify a very successful attack run by an expert attacker with major impact, due to big restore costs.

The success rate is directly proportional with attacker's profile score and the restore cost and decreases with the protection degree, as presented in figure 14.

The relation between system protection degree and attack success rate, for some attacker profiles or attack scenarios will be discussed in a future work.

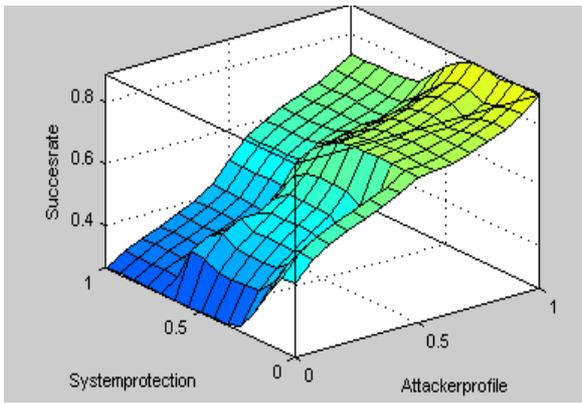

Figure 14. The case of medium system vulnerabilities and big restore costs

## V. CONCLUSIONS

The control systems security is a challenging research topic. Modelling and estimating the attacker types and attack success rate is a concern of the academic and industrial communities. In this paper the authors use a fuzzy inference approach to introduce a new concept – the attacker profile score.

This score indicate how skilled is an attacker based on three attributes: knowledge, technical resources and motivation. Taking into account the authors practical experience in the security field and the data from literature some well-established attacker profiles were analyzed.

In the last part of the paper a fuzzy inference system for assessing the attack success rate is presented. This rate is influenced by four parameters: the attacker profile score previously established, the system protection level, the system vulnerabilities and the restore cost. It shows clearly that the success rate is directly proportional with attacker's profile score and the restore cost and decreases with the protection degree.

Further research should be done in estimating the security risks associated of a specific control system tacking into account the attacker profile score and the attack success rate defined in this paper. There is also needed a context-aware approach of security, that take into consideration the changes in time of motivation and vulnerabilities in order to implement dynamic countermeasures in control system, without affecting their performances and objectives.